\documentclass{article}
\usepackage[T1]{fontenc} 
\usepackage[utf8]{inputenc} 
\usepackage{ismir,amsmath,cite,url}
\usepackage{graphicx}
\usepackage{color}
\usepackage{multirow}
\usepackage{booktabs}
\usepackage{lineno}
\usepackage{subfigure}
\usepackage{amsfonts}

\title{AEROMamba: An efficient architecture for audio super-resolution using generative adversarial networks and state space models
}

\oneauthor
  {Wallace Abreu, Luiz W. P. Biscainho} {Federal University of Rio de Janeiro \\ {\tt \{wallace.abreu, wagner\}@smt.ufrj.br}}

\def\authorname{W. Abreu and L. W. P. Biscainho}

\begin{document}

\maketitle
\begin{abstract}
Audio super-resolution aims to enhance low-resolution signals by creating high-frequency content. In this work, we modify the architecture of AERO (a state-of-the-art system for this task) for music super-resolution. SPecifically, we replace its original Attention and LSTM layers with Mamba, a State Space Model (SSM), across all network layers. Mamba is capable of effectively substituting the mentioned modules, as it offers a mechanism similar to that of Attention while also functioning as a recurrent network. With the proposed AEROMamba, training requires 2-4x less GPU memory, since Mamba exploits the convolutional formulation and leverages GPU memory hierarchy. Additionally, during inference, Mamba operates in constant memory due to recurrence, avoiding memory growth associated with Attention. This results in a 14x speed improvement using 5x less GPU. Subjective listening tests (0 to 100 scale) show that the proposed model surpasses the AERO model. In the MUSDB dataset, degraded signals scored 38.22, while AERO and AEROMamba scored 60.03 and 66.74, respectively. For the PianoEval dataset, scores were 72.92 for degraded signals, 76.89 for AERO, and 84.41 for AEROMamba.

\end{abstract}
\section{Introduction}\label{sec:introduction}

Audio super-resolution is a technique analogous to what is known in signal processing literature as bandwidth extension~\cite{Larsen04}, whose goal is to reconstruct the upper spectral content of a low-resolution signal. Since a bandlimited audio signal usually sounds muffled, higher-resolution audio yields a better listening experience, in general~\cite{Larsen04}.

Since the 19th-century invention of sound recording devices~\cite{IASA14}, audio signals have been widely used in communications and entertainment. Technology has evolved to meet specific application requirements, with telephony prioritizing intelligibility and general audio devices focusing on fidelity~\cite{Biscainho17}. High-fidelity systems must cover at least the human auditory range of 20 Hz to 20 kHz for tones~\cite{Bosi02}, though analog audio may face limitations and media degradation affecting this content~\cite{Copeland08}. In digital audio, compression lowers transmission and storage costs. Decimation, which discards samples, needs low-pass filtering to prevent aliasing and reduce the signal's maximum frequency. Lossy audio coding, such as MP3~\cite{Brandenburg99}, modifies frequency content based on a psychoacoustic model, mostly affecting high frequencies and sometimes reducing bandwidth. Audio super-resolution is useful in scenarios requiring mitigation of these issues.

Signal processing-based bandwidth extension methods included techniques such as nonlinear devices with linear filtering~\cite{Larsen04}, source-filter modeling~\cite{Spanias06}, codebook mapping~\cite{Schmidt08}, and spectral-band replication~\cite{Ekstrand02}. In the past few years, solutions based on deep neural networks (DNNs) became the state of the art in audio super-resolution, ranging from pure feedforward networks that operate on raw waveforms~\cite{Kuleshov17} or in the spectral domain~\cite{Lagrange20} to generative solutions using Generative Adversarial Networks (GANs)~\cite{Valimaki22, Mandel22} and, more recently, Diffusion Models (DMs)~\cite{Moliner22, Chen24, Moliner2024, Moliner2024BABE2}.

The choice of using DMs instead of GANs is generally justified by training instabilities, suboptimal
mode coverage, and lack of explainability of GANs, while DMs are modeled by statistical physics~\cite{Ho2020DDPM}. Even though efficient DMs~\cite{Ulhaq2024}
constitute an active area of research, their Markov Chain-based sampling requires
sequential inference, which makes parallelization difficult and sample generation
slower compared to GANs. Additionally, narrow mode coverage is only problematic
when diverse data generation is needed, which may not necessarily be the case for
audio enhancement.

In this context, this work proposes improvements to the state-of-the-art AERO~\cite{Mandel22} architecture for super-resolution by replacing its Attention and LSTM layers with Mamba~\cite{Gu2024Mamba}, a State Space Model (SSM) created for efficient sequence modeling which has shown promising results when used in speech enhancement~\cite{Chao24}.

The advantages of this approach are significant: training requires 2x to 4x less GPU memory, and inference runs with a 14x speed gain using 5x less GPU, all with an increase in audio quality, as demonstrated by the listening tests performed. 

\section{Method}\label{sec:method}
AERO is an audio super-resolution GAN that processes music and speech signals, inspired by Demucs~\cite{Defossez22}. Its architecture is composed of a generator based on the U-Net, with Attention and BiLSTMs on its last two encoder blocks, and a MelGAN~\cite{Kumar19} multi-scale discriminator. 

\begin{figure}[!htp]
    \centering
    {\includegraphics[width=0.33\textwidth]{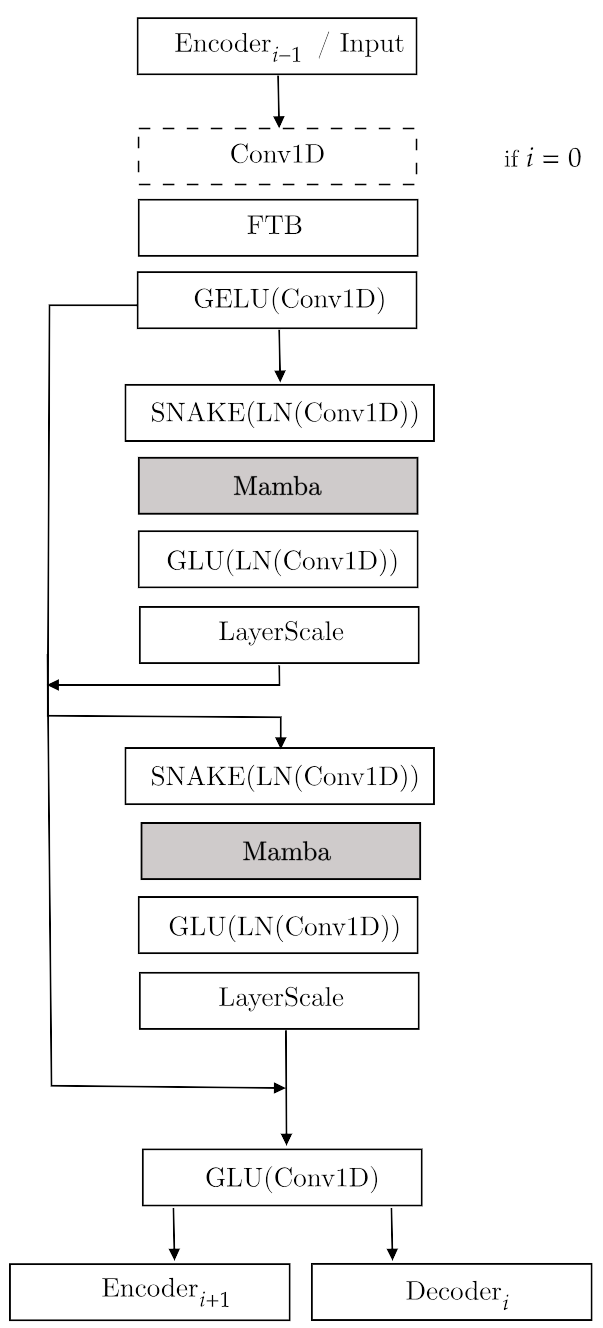}} 
    \caption{Detail of the AEROMamba architecture: AERO with BiLSTMs and Local Attentions replaced by Mamba layers in all Encoder blocks.}
    \label{fig:aeromamba}
\end{figure}

In this work, we propose \textbf{AEROMamba}, in which we replace all the Local Attention and BiLSTM layers in AERO by Mamba layers, including them in all encoder blocks, not only at specific depths. This modification is motivated by our hypothesis that Mamba can handle both tasks effectively, since it offers a selection mechanism similar to Attention while being also a recurrent network, as a generalized SSM. Specifically, this selection mechanism works through a parameterization of the SSM matrices in relation to the input, different from the linear time-invariant formulation. Additionaly, since Mamba exploits the GPU memory hierarchy to perform computations efficiently, we aim to yield high-quality samples while also reducing the use of computational resources in training and inference. The resulting architecture is shown in~\figref{fig:aeromamba}.

Consistently with the original model, we optimize the generator loss
\begin{equation}\label{eq:aero_g_loss}
    L_{\mathcal{G}} = L_\text{adv} + L_\text{rec} + \lambda L_{\text{fmap}},
\end{equation}
composed of the adversarial loss $L_\mathrm{adv}$, the reconstruction loss $L_\mathrm{rec}$, and the features loss $L_\mathrm{fmap}$ with $\lambda=100$. In addition, the training objective $L_{\mathcal{D}}$ for the discriminator is the MelGAN hinge loss~\cite{Kumar19}.

\section{Experiments}\label{sec:experiments}

\subsection{Datasets}
The PianoEval data set\footnote{The metadata of each recording and code repository are available in the accompanying website \url{https://aeromamba-super-resolution.github.io/}.} was compiled in two segments: the first segment (training and validation sets) comprises 45 recordings of the Chopin's 24 Preludes, Op. 28, played by 33 different pianists (totaling 22 hours), while the second segment (test set) incorporates excerpts taken from piano pieces by Ligeti, Schumann and Barber, performed respectively by 3 different performers (amounting to 3.5 hours).  We also tested our model on MUSDB18~\cite{musdb18-hq}, which contains 150 songs (10 hours) of musical mixtures with their
isolated stems, using only the mixture tracks. All files are in WAV format, stereo, sampled at 44.1 kHz.

\subsection{Training procedure}
Originally, AERO was trained in the upsampling configuration from 11.025 kHz to 44.1 kHz using a specific window size $W=512$ along with three distinct hop lengths: $H=64$, $H=128$, and $H=256$. Our model also works in the 11.025-44.1 kHz setting but, since $H=64$ or $H=128$ settings took impractical training time, our development concentrated on the $H=256$ version.

For training on MUSDB18, we started with a pre-trained AERO model (already trained for 696 epochs on the same dataset) and extended its training for an additional 100 epochs, selecting the optimal model based on the Perceptual Audio Quality Measure (PAQM)~\cite{Beerends1992} score. To ensure consistency, AEROMamba was trained for an equivalent number of epochs. The original train/test partition of MUSDB18 was retained.

In the case of PianoEval training set, we created two groups: PianoEval-GQ and PianoEval-HQ (with a difference of 1.5 hours between them) --- the first containing all recordings, and the second contained only tracks recorded after the 1960s. The motivation for this was to evaluate whether the greater presence of noise, intrinsic to older recordings, would be beneficial, adding robustness to the models, or would harm the quality of the predicted results. Both AERO and AEROMamba were trained for approximately 800 epochs, also saving the best model. 

All other hyperparameters were configured according to the default settings of AERO, such as the learning rate of $3\mathrm{e-}4$. Both models underwent training on an NVIDIA GeForce RTX 3090 GPU with the maximum batch size allowed for each model.

\subsection{Evaluation}
For the objective evaluation, we employed the Log-Spectral Distance (LSD)~\cite{Mandel22} and two distinct perceptual quality metrics: PAQM and the Virtual Speech Quality Objective Listener (ViSQOL)~\cite{Hines2015}. Furthermore, we conducted subjective listening tests to provide qualitative comparisons between AERO and AEROMamba on the MUSDB18 and PianoEval datasets. 

PAQM was used as a validation metric to select the optimal model during a training run, with fixed seed and hyperparameters. The choice of PAQM as a validation metric was motivated by the availability of a vectorized implementation, which is much faster than ViSQOL, and also to its perceptual motivation, taking into account masking effects and other auditory modeling aspects~\cite{Beerends1992}. In addition, ViSQOL (in audio mode) was employed to assess the quality of the processed signals on a scale from 1 to 5, in relation to the test sets.

For the listening tests, the opinions of 20 subjects were compiled regarding the overall similarity of three test signals, one corresponding to each model and one anchor, in relation to a known reference. The score was given through a sliding bar without explicit numerical values, with the words `Exactly the same' on the right end and increasing difference indicated by a left arrow. Although the numerical value was not explicit to the subjects, the scale was defined from 0 to 100, with a 1-point resolution. A total of 12 tracks from the PianoEval dataset and 12 tracks from the MUSDB18-HQ test set were employed, with each track being evaluated by 10 subjects. The order of questions was randomized to mitigate bias.

Test signals were selected to ensure maximum variety in terms of dynamic range, tempo, spectral content, and sound intensity. In the case of PianoEval, this was achieved by selecting four different excerpts from each composer. For MUSDB18HQ, two pieces were selected for each genre, namely rock, pop, electronic, hip hop, world music, and other (an additional pair of songs with no specified genre). The duration of the signals was set between 15 and 20 seconds, according to the standards~\cite{ITUBS1116} and providing sufficient listening content for the subjects, with fade-in and fade-out effects to smooth abrupt starts or ends when needed.

\section{Results}\label{sec:results}
Results and computational performance metrics are summarized in Tables~\ref{tab:musdb18_visqol_lsd_scores_comparison},~\ref{tab:pianoeval_visqol_lsd_scores_comparison}, and~\ref{tab:performance_comparison}, which make it clear that AEROMamba (with its corresponding HQ version) achieved superior performance to AERO objectively and subjectively, using almost 6x-9x less GPU in inference and running almost 15x as fast (in addition to training at least 2x faster). These performance improvements are due to the efficiency provided by the Mamba layer and also to its ability in sequence modeling tasks. As seen in Table~\ref{tab:performance_comparison}, we were able to build a larger architecture, in the sense of parameters, that uses fewer computational resources, thus theoretically being a more powerful model. 
\begin{table}[!ht]
\centering
\begin{tabular}{lccc}
\toprule
\multirow{2}{*}{Model} & \multicolumn{3}{c}{MUSDB18} \\
\cmidrule(lr){2-4}
 & ViSQOL $\uparrow$ & LSD $\downarrow$ & Score $\uparrow$ \\
\midrule
Low-Resolution      & 1.82 & 3.98 & 38.22 \\
AERO                & 2.90 & 1.34 & 60.03 \\
AEROMamba           & \textbf{2.93} & \textbf{1.23} & \textbf{66.47} \\
\bottomrule
\end{tabular}
\caption{ViSQOL, LSD, and subjective scores for AERO and AEROMamba on the MUSDB18 dataset.}
\label{tab:musdb18_visqol_lsd_scores_comparison}
\end{table}
\begin{table}[!ht]
\centering
\begin{tabular}{lccc}
\toprule
\multirow{2}{*}{Model} & \multicolumn{3}{c}{PianoEval} \\
\cmidrule(lr){2-4}
 & ViSQOL $\uparrow$ & LSD $\downarrow$ & Score $\uparrow$ \\
\midrule
Low-Resolution      & 4.36 & 1.09 & 72.92 \\
AERO                & \textbf{4.38} & \textbf{0.99} & 76.89 \\
AERO-HQ             & 4.34 & 1.04 & -     \\
AEROMamba           & 4.43 & 0.98 & -     \\
AEROMamba-HQ        & \textbf{4.38} & 1.00 & \textbf{84.41} \\
\bottomrule
\end{tabular}
\caption{ViSQOL, LSD, and subjective scores for AERO and AEROMamba on the PianoEval dataset. Models labeled with `-HQ` were trained on PianoEval-HQ. }
\label{tab:pianoeval_visqol_lsd_scores_comparison}
\end{table}
\begin{table*}[!ht]
\centering
\begin{tabular}{lccccp{3cm}}
\toprule
\multirow{2}{*}{Method} & \multicolumn{2}{c}{NVIDIA RTX 3090} & \multicolumn{2}{c}{NVIDIA RTX 2080 Ti} & \multirow{2}{*}{Parameters} \\
\cmidrule(lr){2-3} \cmidrule(lr){4-5}
                        & GPU Usage (MB) & Time (s) & GPU Usage (MB) & Time (s) & \\
\midrule
AERO                    & 17091          & 1.246    & 16420*          & --       & 19,432,958 \\
AEROMamba               & 3000           & 0.087    & 1914           & 0.063    & 20,964,190 \\
\bottomrule
\end{tabular}%
\caption{Comparison of GPU usage, inference times for 10 second segments, and number of parameters for AERO and AEROMamba. *AERO exceeded the GPU memory limit by 5.16 GB.}
\label{tab:performance_comparison}
\end{table*}

According to Mann-Whitney~\cite{Gibbons85} tests on PianoEval ViSQOL results, all models when paired with Low-Resolution scores distribution achieved statistical significance with $p$-value < 0.05, except for AeroMamba. Therefore, we decided to evaluate AEROMamba-HQ in subjective tests, together with AERO, due to its higher ViSQOL score compared to its `-HQ' version. For subjective scores, all pairs were considered statistically different. Considering that PianoEval-HQ is 1.5 hours shorter in duration than PianoEval, we demonstrate a scenario where AEROMamba architecture can attain scores superior to AERO's even with a reduced amount of data. 

Regarding the results of MUSDB18-HQ, the same Mann-Whitney tests reported statistical significance with $p$-value < 0.05 between all distribution pairs, both for ViSQOL and subjective listening tests scores. 

\begin{figure}[!h]
    \centering
    \subfigure[]{\includegraphics[width=0.47\textwidth]{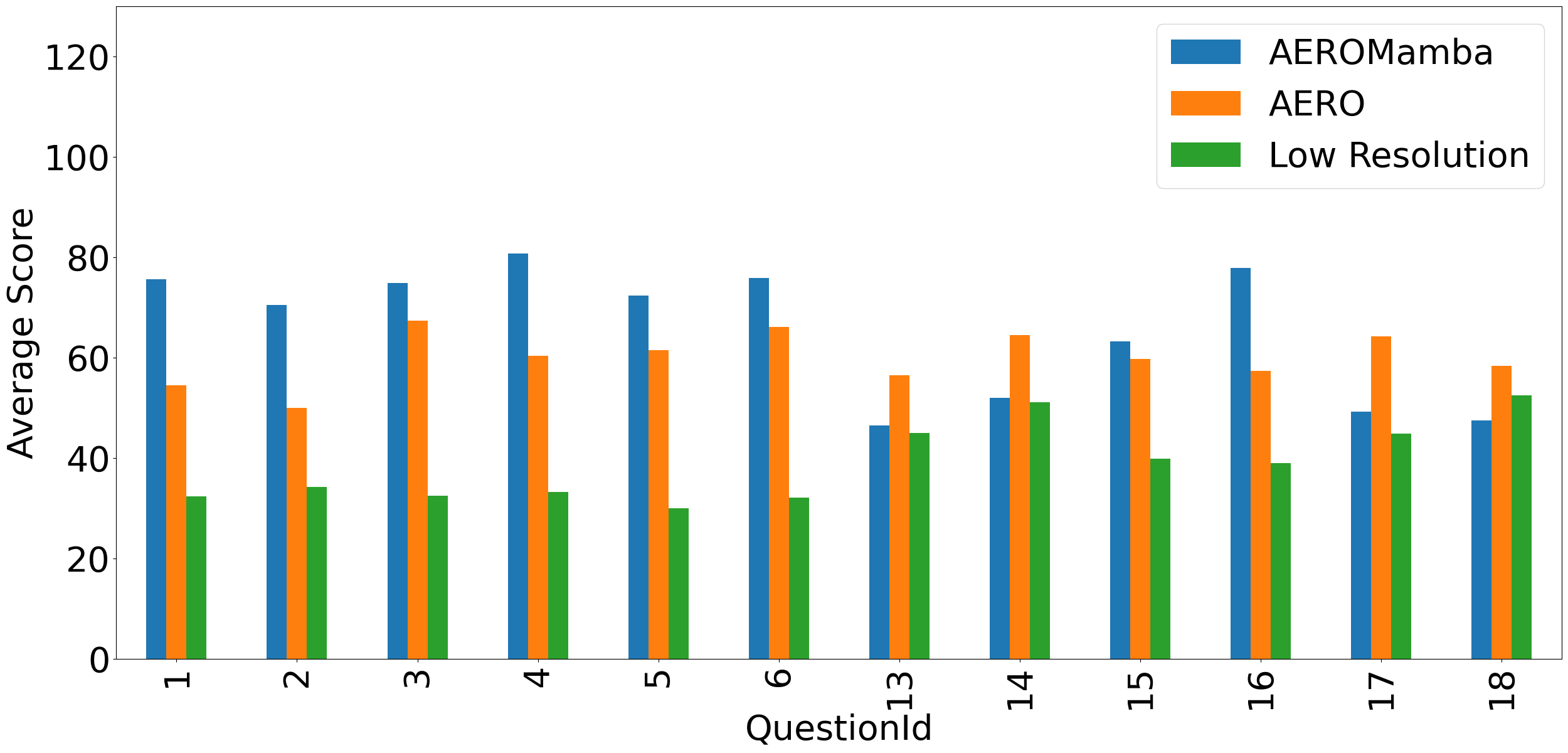}} 
    \subfigure[]{\includegraphics[width=0.47\textwidth]{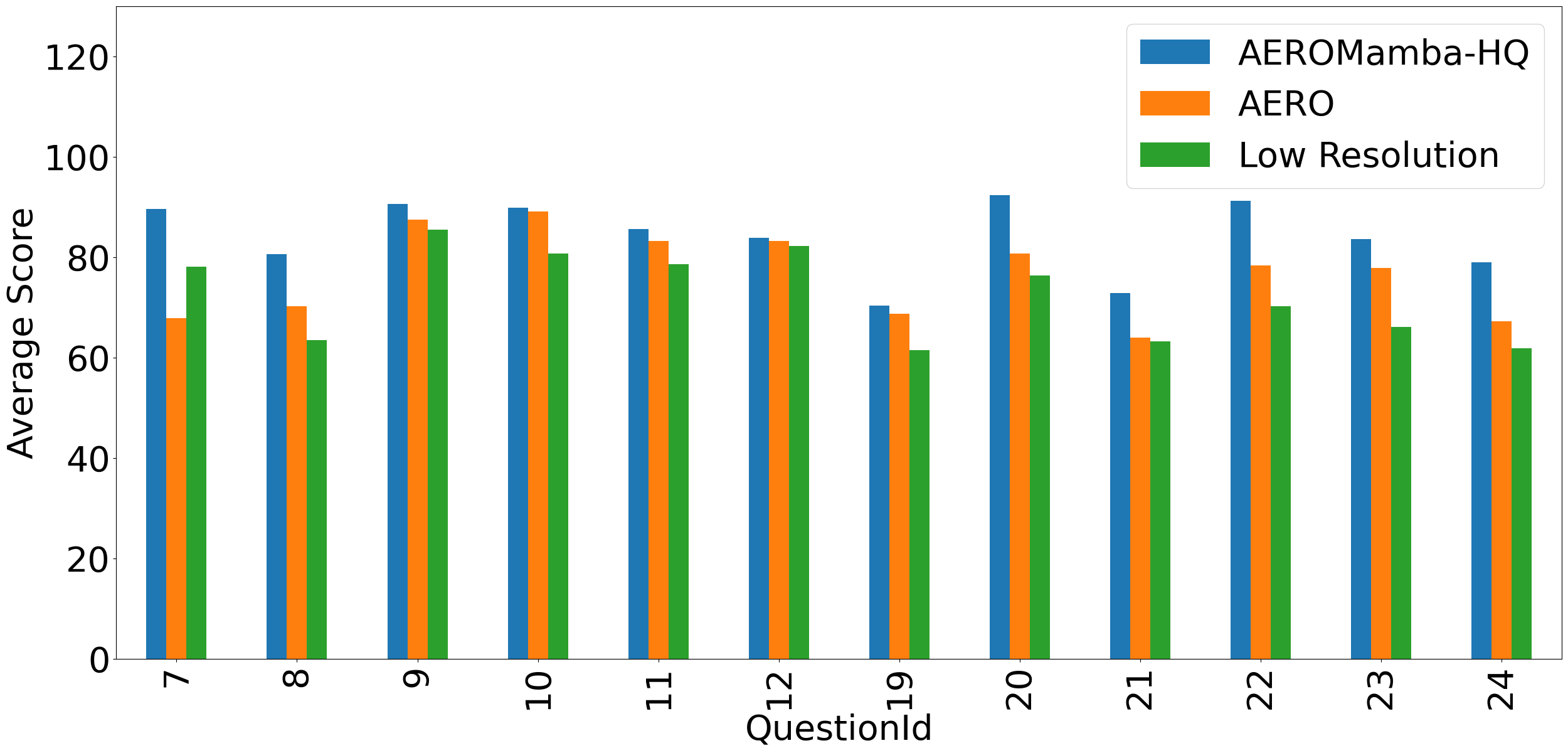}} 

    \caption{Subjective scores per track id obtained (a) on MUSDB18; (b) on PianoEval.}
    \label{fig:subjective_scores}
\end{figure}

For a detailed visualization, we show in~\figref{fig:subjective_scores} the scores for each track included in the subjective evaluation procedure, identified by their Id number on the testing interface. It is clear that the negative effect of downsampling is more severe on the MUSDB18 tracks, while for PianoEval tracks the improvement of super-resolution is less pronounced. This is explained by the nature of the two datasets: while PianoEval content is limited to a single acoustic instrument played with varied dynamics and agogics, MUSDB18-HQ tracks contain a wide range of electronic sounds and percussion, usually in a heavily saturated mix and high tempo. 

\begin{figure}[!h]
    \centering
    \subfigure[]
    {\includegraphics[width=0.42\textwidth]{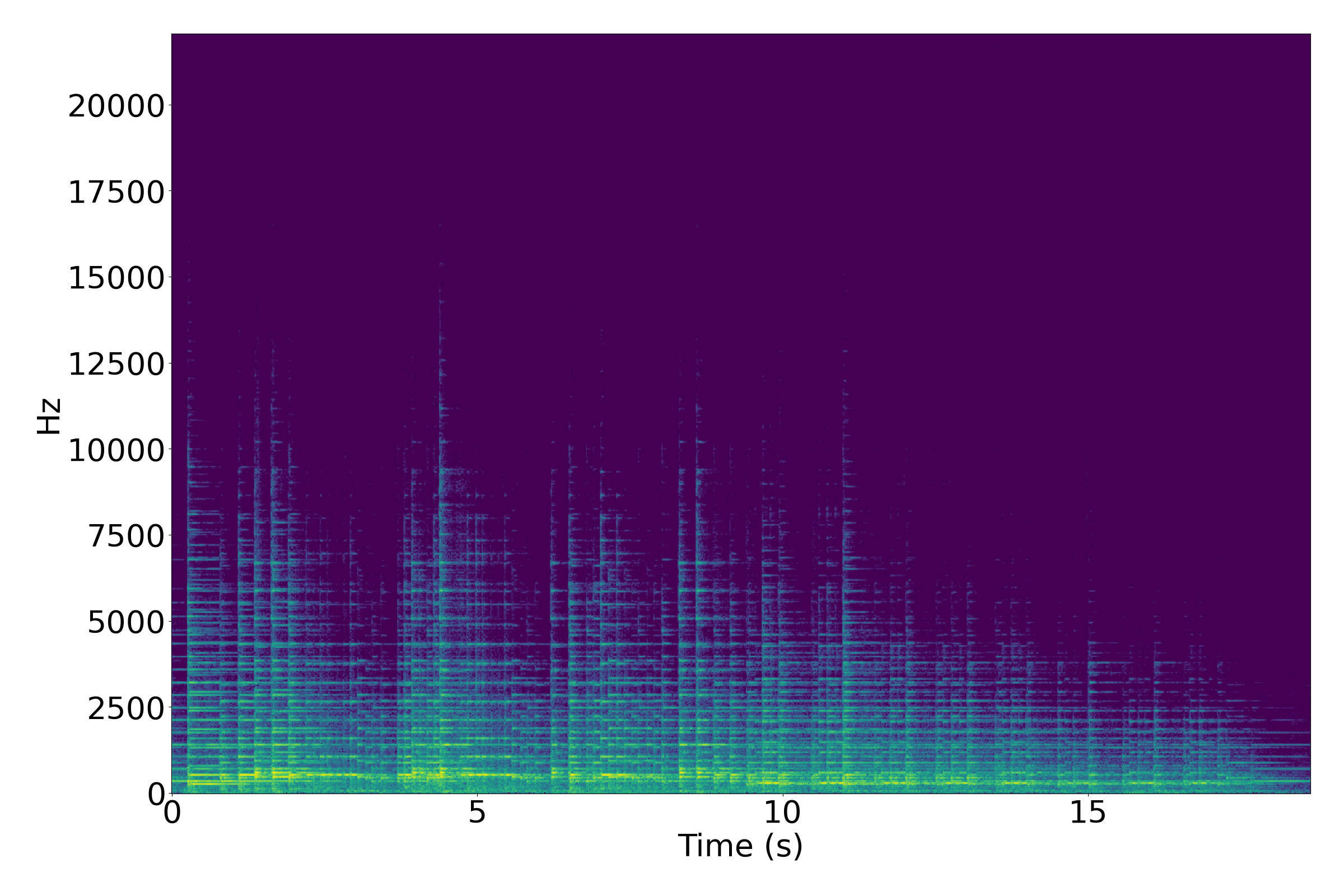}} 
    \subfigure[]
    {\includegraphics[width=0.42\textwidth]{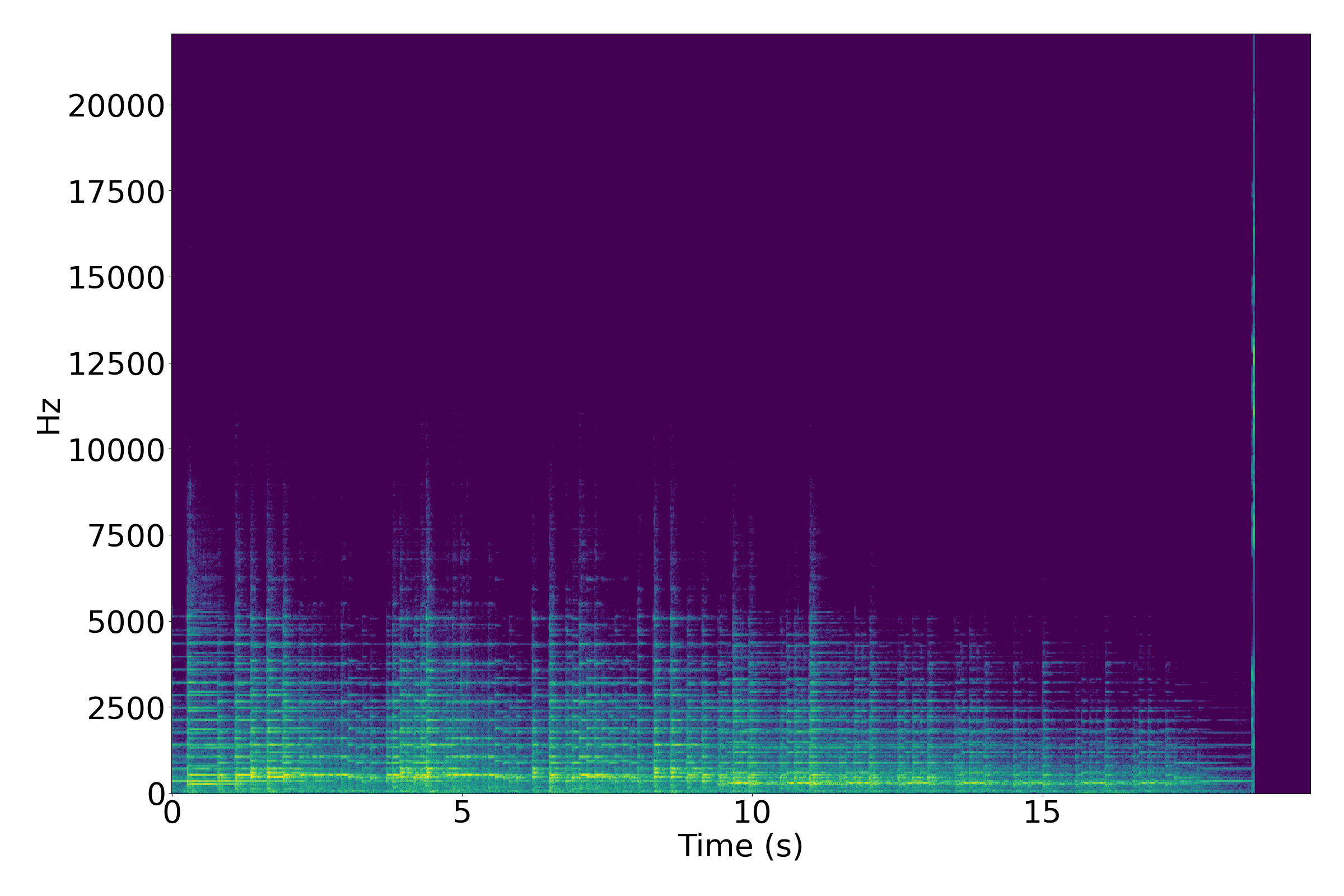}} 
    \subfigure[]{\includegraphics[width=0.42\textwidth]{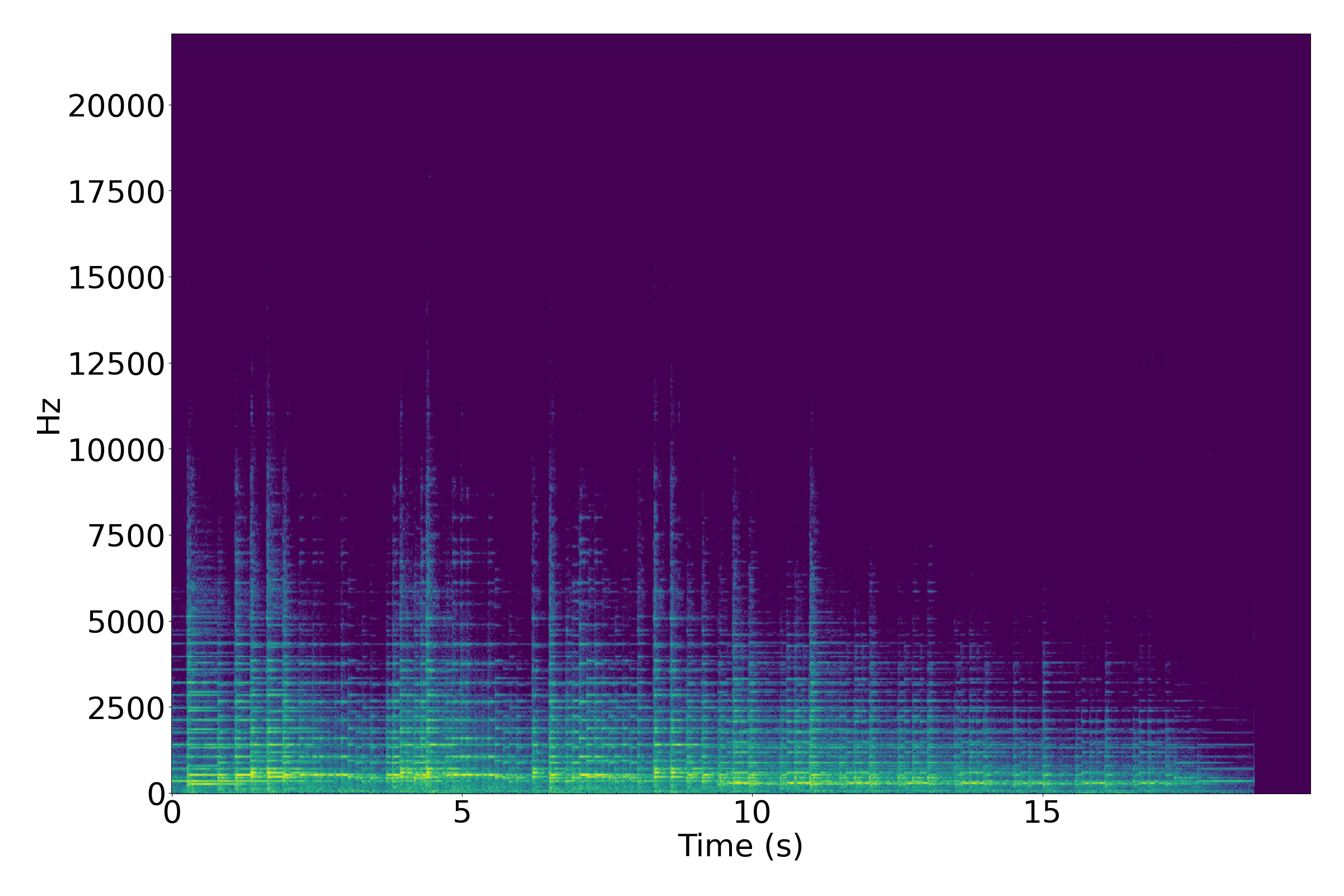}} 

    \caption{Frequency spectra of Track 22 of PianoEval: (a) original; (b) processed by AERO. (c) processed by AEROMamba-HQ.}
    \label{fig:piano_specs}
\end{figure}

As a qualitative illustration, we compare in~\figref{fig:piano_specs} the frequency spectra of the enhanced signals for AERO and AEROMamba-HQ, using Track 22 of PianoEval as a reference. Evidently, most of the signal's energy is below 5 kHz, which explains why its low-resolution version scored above 70. However, there is also clear differences between the results of two models. AEROMamba produces a greater amount of high-frequency content than AERO. No visible artifacts are introduced, except for an impulse introduced by AERO at the end of the track due to the abrupt silence induced by zero-padding (easily avoidable).

Finally, in the case of PianoEval, listeners reported that AEROMamba was capable of creating more high-frequency content than AERO, generating a brighter sound that seemed closer to the original, as discussed above. However, this same behaviour resulted in the perception of unexpected artifacts in MUSDB18 evaluation, which led AERO to be indicated as the best model in four tracks.

\section{Conclusion}\label{sec:conclusion}
In this work we proposed an efficient neural network architecture based on a state-of-the-art solution for audio super-resolution. Our method significantly reduces GPU usage and offers faster inference without compromising audio quality. We confirm the superiority of our model to the baseline both through objective metrics and by evaluating the subjective quality of our model via listening tests.

For future works, one of the main modifications that can benefit the usability of the model is to implement a flexible initial sampling frequency, instead of just 11.025 kHz. In addition, we would like to evaluate whether the models are invariant to similar instruments, such as the piano and the harpsichord, to the point of achieving good performance on one when trained on the other.

\section{Acknowledgments}
The authors thank CNPq, FAPERJ, and CAPES for sponsoring this research.

\bibliography{LAMIR2024_template}

%
%
%
%
%

\end{document}